\documentclass[aps,prl,preprint,tightenlines,superscriptaddress,showpacs,byrevtex]{revtex4}
\usepackage{graphicx} 
\usepackage{dcolumn}  

\graphicspath{{ps}}

\begin{document}


\preprint{\vbox{ \hbox{   }
                 \hbox{BELLE-CONF-0509}
                 \hbox{LP2005-144}
                 \hbox{EPS05-481} 
}}

\title{ \quad\\[0.5cm]  Measurement of the mass of the $\tau$-lepton
and an upper limit on the
mass difference between $\tau^+$ and $\tau^-$ 
}

\affiliation{Aomori University, Aomori}
\affiliation{Budker Institute of Nuclear Physics, Novosibirsk}
\affiliation{Chiba University, Chiba}
\affiliation{Chonnam National University, Kwangju}
\affiliation{University of Cincinnati, Cincinnati, Ohio 45221}
\affiliation{University of Frankfurt, Frankfurt}
\affiliation{Gyeongsang National University, Chinju}
\affiliation{University of Hawaii, Honolulu, Hawaii 96822}
\affiliation{High Energy Accelerator Research Organization (KEK), Tsukuba}
\affiliation{Hiroshima Institute of Technology, Hiroshima}
\affiliation{Institute of High Energy Physics, Chinese Academy of Sciences, Beijing}
\affiliation{Institute of High Energy Physics, Protvino}
\affiliation{Institute of High Energy Physics, Vienna}
\affiliation{Institute for Theoretical and Experimental Physics, Moscow}
\affiliation{J. Stefan Institute, Ljubljana}
\affiliation{Kanagawa University, Yokohama}
\affiliation{Korea University, Seoul}
\affiliation{Kyoto University, Kyoto}
\affiliation{Kyungpook National University, Taegu}
\affiliation{Swiss Federal Institute of Technology of Lausanne, EPFL, Lausanne}
\affiliation{University of Ljubljana, Ljubljana}
\affiliation{University of Maribor, Maribor}
\affiliation{University of Melbourne, Victoria}
\affiliation{Nagoya University, Nagoya}
\affiliation{Nara Women's University, Nara}
\affiliation{National Central University, Chung-li}
\affiliation{National Kaohsiung Normal University, Kaohsiung}
\affiliation{National United University, Miao Li}
\affiliation{Department of Physics, National Taiwan University, Taipei}
\affiliation{H. Niewodniczanski Institute of Nuclear Physics, Krakow}
\affiliation{Nippon Dental University, Niigata}
\affiliation{Niigata University, Niigata}
\affiliation{Nova Gorica Polytechnic, Nova Gorica}
\affiliation{Osaka City University, Osaka}
\affiliation{Osaka University, Osaka}
\affiliation{Panjab University, Chandigarh}
\affiliation{Peking University, Beijing}
\affiliation{Princeton University, Princeton, New Jersey 08544}
\affiliation{RIKEN BNL Research Center, Upton, New York 11973}
\affiliation{Saga University, Saga}
\affiliation{University of Science and Technology of China, Hefei}
\affiliation{Seoul National University, Seoul}
\affiliation{Shinshu University, Nagano}
\affiliation{Sungkyunkwan University, Suwon}
\affiliation{University of Sydney, Sydney NSW}
\affiliation{Tata Institute of Fundamental Research, Bombay}
\affiliation{Toho University, Funabashi}
\affiliation{Tohoku Gakuin University, Tagajo}
\affiliation{Tohoku University, Sendai}
\affiliation{Department of Physics, University of Tokyo, Tokyo}
\affiliation{Tokyo Institute of Technology, Tokyo}
\affiliation{Tokyo Metropolitan University, Tokyo}
\affiliation{Tokyo University of Agriculture and Technology, Tokyo}
\affiliation{Toyama National College of Maritime Technology, Toyama}
\affiliation{University of Tsukuba, Tsukuba}
\affiliation{Utkal University, Bhubaneswer}
\affiliation{Virginia Polytechnic Institute and State University, Blacksburg, Virginia 24061}
\affiliation{Yonsei University, Seoul}
  \author{K.~Abe}\affiliation{High Energy Accelerator Research Organization (KEK), Tsukuba} 
  \author{K.~Abe}\affiliation{Tohoku Gakuin University, Tagajo} 
  \author{I.~Adachi}\affiliation{High Energy Accelerator Research Organization (KEK), Tsukuba} 
  \author{H.~Aihara}\affiliation{Department of Physics, University of Tokyo, Tokyo} 
  \author{K.~Aoki}\affiliation{Nagoya University, Nagoya} 
  \author{K.~Arinstein}\affiliation{Budker Institute of Nuclear Physics, Novosibirsk} 
  \author{Y.~Asano}\affiliation{University of Tsukuba, Tsukuba} 
  \author{T.~Aso}\affiliation{Toyama National College of Maritime Technology, Toyama} 
  \author{V.~Aulchenko}\affiliation{Budker Institute of Nuclear Physics, Novosibirsk} 
  \author{T.~Aushev}\affiliation{Institute for Theoretical and Experimental Physics, Moscow} 
% ...
\collaboration{The Belle Collaboration}

\noaffiliation

\begin{abstract}
The mass of the $\tau$-lepton has been measured in the decay modes
$\tau \rightarrow 3\pi \nu_\tau$ and $\tau \rightarrow 3\pi \pi^0\nu_\tau$
using a pseudomass technique. The preliminary result is 
$1776.71\pm 0.25 \mbox{(stat)} \pm 0.62 \mbox{(syst)}$ MeV.
The preliminary value of an upper limit on the relative mass difference 
between positive and negative $\tau$ leptons is
$|(M_{\tau^+}-M_{\tau^-})|/M_{\mathrm{average}}$ is $5.0 \times 10^{-4}$
at 90\% CL.
\end{abstract}


\maketitle

\tighten

{\renewcommand{\thefootnote}{\fnsymbol{footnote}}}
\setcounter{footnote}{0}
\section{Introduction}
 Masses of quarks and leptons are fundamental parameters of the
Standard Model (SM).
In the SM high precision measurements of the mass, lifetime and
leptonic branching fractions of the $\tau$ lepton can be used to test 
lepton universality. The present PDG value of the 
$\tau$ mass~\cite{PDG} is dominated by the result of the BES 
Collaboration~\cite{BES} and has 
an accuracy about 0.3 MeV. The high statistics of the Belle
data allow a measurement with the same level
of accuracy. The methods used for the $\tau$ mass
measurement are different for the BES and the Belle experiments:
BES analysed the cross section for $\tau$ pair production near threshold
while in Belle the four-momenta of the visible $\tau$ decay products 
at $\sqrt{s}$=10.58 GeV are used.
This leads to different sources of systematic uncertainties.
Eventually, by combining these two measurements we could
significantly improve the accuracy of the $\tau$ mass deterination.

The analysis of individual $\tau$ lepton decays allows to measure
the masses of positive and negative $\tau$'s separately and  
test the CPT theorem. 
A similar test was performed by OPAL at LEP~\cite{OPAL} with the result
$(M_{\tau^+}-M_{\tau^-})/M_{\mathrm{average}}<3.0\times 10^{-3}$ at 90\% CL.
The high statistics and quality of the Belle data allow us to improve 
this limit significantly. 

To determine the $\tau$ mass we use a pseudomass technique that
was first employed by the ARGUS~\cite{ARGUS} collaboration. This technique relies
on the reconstruction of the invariant mass and energy of the hadronic system 
in hadronic $\tau$ decays.

\section{The pseudomass method}
In a hadronic $\tau$ decay (see Fig. 1) the $\tau$ lepton mass
$M_\tau$ is related to the four-momentum of the resulting hadronic system X
by the formula
 
\begin{equation}
 M_\tau^2=M_X^2+M_\nu^2+2E_XE_\nu-2P_XP_\nu \cos\theta
\end{equation}

\noindent where $M_X$, $E_X$ and $P_X$ are the invariant mass, 
energy and absolute
value of the momentum of the hadronic system respectively; $M_\nu$, $E_\nu$ 
and $P_\nu$ are the same quantities for the neutrino and $\theta$ is
the angle between the momenta of the neutrino and hadronic system.

\begin{figure}[htb]
\includegraphics[width=0.54\textwidth]{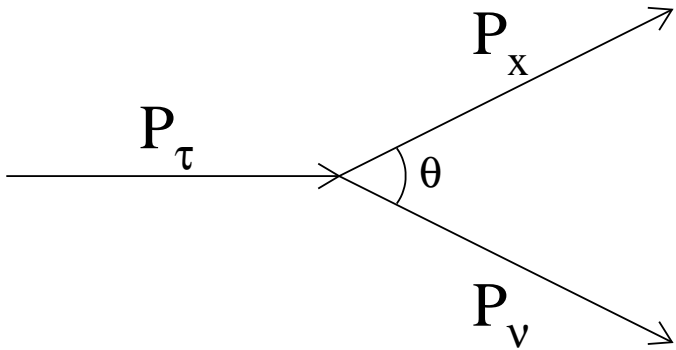}
\caption{Illustration for the variable definitions used in the Eq.(1)}
\label{fg01}
\end{figure}

If $M_\nu=0$ we have $P_\nu=E_\nu = E_\tau-E_X$. 
This gives the following expression for the $\tau$ mass:


\begin{equation}
 M_\tau^2=M_X^2+2(E_\tau-E_X)(E_X-P_X \cos\theta).
\end{equation}

The $\tau$ lepton energy $E_\tau$ is obtained from
the energy of the electron or positron beam, $E_{\mathrm{beam}}$,
in the center-of-mass (c.m.) frame. All other kinematic variables
listed above will also be evaluated in the c.m. frame 
of the colliding beams.


 If we set the unknown value of $\cos\theta$ in equation (2) equal
to 1, the right side of (2) will be smaller than the
true value for $M_\tau^2$. Therefore, the estimator of the $\tau$ mass
(the so-called pseudomass) used in the analysis

\begin{figure}[htb]
\includegraphics[width=0.54\textwidth]{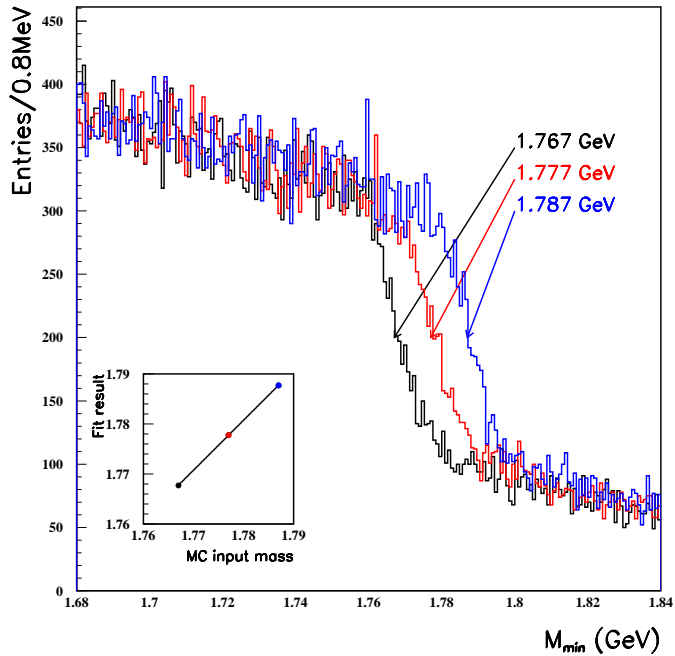}
\caption{The Monte Carlo distributions of the pseudomass $M_{\mathrm{min}}$ for
the $\tau \rightarrow 3\pi^\pm \nu$ decays and the input $\tau$ masses
equal to 1.767, 1.777 and 1.787 GeV, respectively.
The fitted values of the $\tau$ mass versus the input mass used in the
MC simulation are shown together with a straight line fit.}
\label{fg02}
\end{figure}

\begin{equation}
 M_{\mathrm{min}}=\sqrt{M_X^2+2(E_{\mathrm{beam}}-E_X)(E_X-P_X)}
\end{equation}
should be less than or equal to the $\tau$ lepton mass. In the absence
of initial and final state radiation and assuming a perfect measurement of
the four-momentum of the hadronic system, $M_{\mathrm{min}}$ is distributed
below the $\tau$ mass and has an edge at $M_\tau$.
The initial and final state radiation as well as the finite momentum
resolution of the detector smear the shape of the edge for $M_{\mathrm{min}}$ around
$M_\tau$. The contributions from background processes in the selected
$\tau^+\tau^-$ sample have smooth behavior near the $\tau$
mass.
We can use the threshold position obtained from the fit to the experimental
$M_{\mathrm{min}}$ distribution as an estimator of the $\tau$ mass.

To illustrate this method in the Belle environment we
performed simulations of $\tau$ decays into three charged
pions and a neutrino with three different input $\tau$ masses: the nominal 
PDG value 1.777 GeV, 
1.767 GeV and 1.787 GeV\footnote{In this paper the masses, momenta and energies
are given in GeV, i.e. we take c=1}. 
The generated events
were passed through the full Belle detector response simulation and
reconstruction procedures. 
The resulting $M_{\mathrm{min}}$ distributions for the three input $\tau$ masses 
given above are shown in Fig. 2. The inset in
Fig. 2 shows the dependence of the fitted masses obtained from
the $M_{\mathrm{min}}$ distributions on the input masses used in the 
MC simulation.
The result of the fit of this dependence to the linear function 
$f=a_0+a_1\times x$
gives $a_0=(0.1\pm0.2)\times 10^{-2}$ GeV and $a_1=1.000\pm0.001$.

 After obtaining the value of the threshold position from the fit
to the experimental $M_{\mathrm{min}}$ distribution we correct it by the value,
obtained from the Monte Carlo, which is equal to the difference
between the true input value of the $\tau$ mass used in the MC simulation and 
the threshold position obtained from the simulated data. 

\section{Experimental Procedure}
 Analysis presented here is based on the data taken with the Belle
detector at the KEKB asymmetric-energy $e^+e^-$ collider.
The total integrated luminosity used in the analysis is 253 fb$^{-1}$.

 A detailed description of the Belle detector is given
elsewhere~\cite{BELLE}. We mention here only the detector components essential
for the present analysis.

 Charged tracks are reconstructed from hit information in a central drift
chamber (CDC) located in a 1.5 T solenoidal magnetic field. The $z$
axis of the detector and the solenoid are along the positron 
beam direction,
with positrons moving in the $-z$ direction. The CDC measures the
longitudinal and transverse momentum components (along the $z$ axis and in the
$r\phi$ plane, respectively). Track trajectory coordinates near the collision
point are provided by a silicon vertex detector (SVD). Photon detection and
energy measurements are performed with a CsI(Tl) electromagnetic calorimeter
(ECL).
Identification of kaons is based on the information from the time-of-flight
counters (TOF) and silica aerogel Cherenkov counters (ACC). The ACC provides
good separation between kaons and pions or muons at momenta above 1.2 GeV.
The TOF system consists of a barrel of 128 plastic scintillation counters, and
is effective in $K/\pi$ separation mainly for tracks with momentum below 1.2
GeV. The lower energy kaons are also identified using specific ionization 
$(dE/dx)$ measurements in the CDC.
Identification of electrons is made using combined information from ECL, ACC,
TOF and CDC~\cite{LE}.
The magnet return yoke is instrumented to form the $K_L$ and muon detector
(KLM), which detects muon tracks~\cite{LMU} and provides trigger signals.

 The signal events were efficiently triggered by several kinds of track triggers
that require two or more CDC tracks with combinations of TOF hits, ECL
clusters or its energy sum. Here, we do not eliminate any events using the
trigger condition information. The trigger conditions are complementary to
each other for the detection of four-prong events in the present case. We can
expect a high trigger efficiency ($\sim$ 95\%) by combining them.

 We used only on-resonance data because the absolute beam energy calibration is
known for this data sample better than for the off-resonance data
taken 60 MeV below the $\Upsilon$(4S)
($\sqrt{s} = 10.58$ GeV ).


$\tau^+\tau^-$ events were selected where one $\tau$ lepton
decays leptonically into $l\bar{\nu_l}\nu_\tau$. The other $\tau$ lepton
decays into hadronic decay modes 
with 3 charged pions and a neutrino or 3 charged pions with 1 neutral pion
and a neutrino. 

The preselection of events is based on the following criteria:

\begin{itemize}
\item[(a)] Visible reconstructed energy $E_{sum}>0.18\sqrt{s}$;
\item[(b)] Number of well-reconstructed charged tracks greater than 2;
\item[(c)] The sum of the z components of each good track
and good photon momenta is required to satisfy $|P_z|<0.5\sqrt{s}$.
\end{itemize}

 Conditions (a) and (c) are calculated in the c.m. frame.

The criteria for good charged tracks are:

\begin{itemize}
\item $p_T > 100$ MeV;
\item Impact parameters $\Delta r < 2$ cm, $|\Delta z| < 4$ cm.
\end{itemize}

 Good photons are defined as ECL clusters with energy greater than 100 MeV 
that are not associated with charged tracks. 
The angular acceptance for photons is $17^\circ<\theta<150^\circ$.

 The preselection cuts suppress Bhabha, $\mu^+\mu^-$ and two-photon events.

After the preselection the following cuts were applied.

\begin{itemize}
\item Total charge equal to zero;
\item Number of leptons (muons or electrons) equal to one;
\item Number of charged pions equal to three;
\item Number of charged kaons and protons equal to zero;
\item Number of $K_S$'s equal to zero;
\item Number of $\pi^0$'s equal to zero for the 
      $\tau \rightarrow 3\pi \nu_\tau$ decay mode and equal to one for the
      $\tau \rightarrow 3\pi \pi^0 \nu_\tau$ mode.
\end{itemize}

\begin{figure}[htb]
\includegraphics[width=0.5\textwidth]{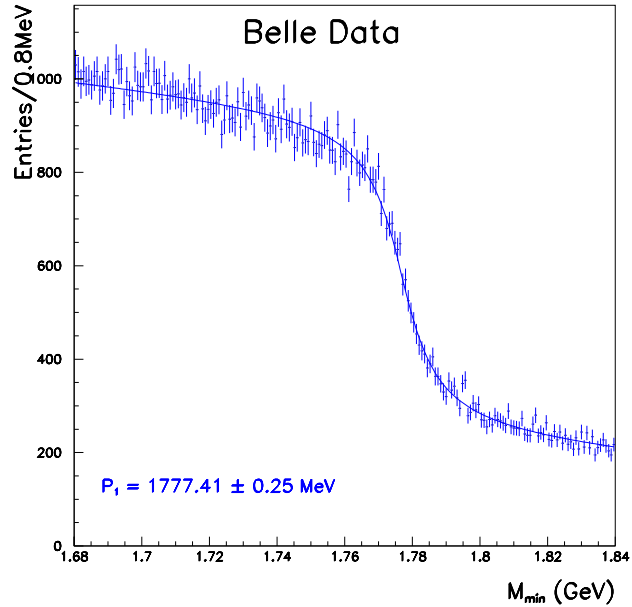}
\caption{The pseudomass distribution $M_{\mathrm{min}}$ for
the $\tau \rightarrow 3\pi^\pm \nu$ decays. The points with error
bars are data and the solid line is the result of the fit with
function (4). }
\label{fg03}
\end{figure}

\begin{figure}[htb]
\includegraphics[width=0.5\textwidth]{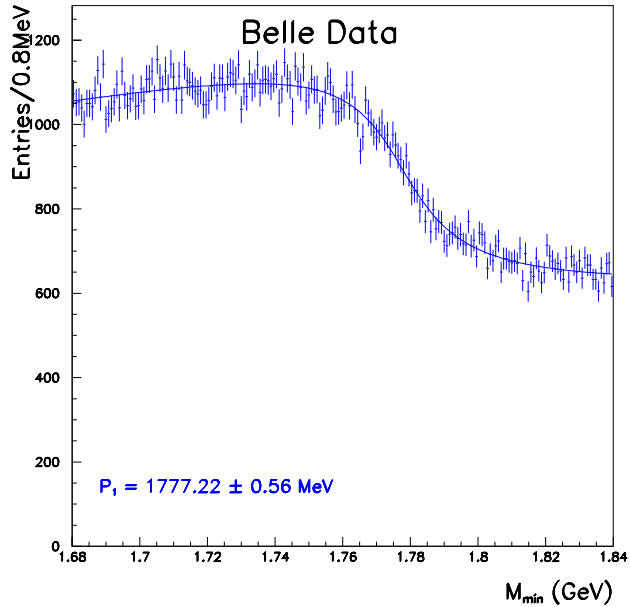}
\caption{The pseudomass distribution $M_{\mathrm{min}}$ for
the $\tau \rightarrow 3\pi^\pm\pi^0 \nu$ decays. The points with error
bars are data and the solid line is the result of the fit with
function (4). }
\label{fg04}
\end{figure}

 The $M_{\mathrm{min}}$ distribution for the 
$\tau \rightarrow 3\pi \nu$ data 
is shown in Fig. 3.
A fit was performed to the data with the function

\begin{equation}
F(x) = (p_3+p_4\times x)\times \mbox{arctan}((x-p_1)/p_2)+p_5+p_6\times x
\end{equation}

\noindent The value of the parameter $p_1$ obtained from the fit is
$p_1 = 1777.41 \pm 0.25$ MeV. 

The difference between the threshold position obtained from using equation
(4) and the true value of the $\tau$ mass obtained from MC is equal to   
$\delta p_1 = 0.70\pm0.40$ MeV. 
The uncertainty in $\delta p_1$ is dominantly due to
limited Monte Carlo statistics (which is about 1/2
of the data) and the systematics of the fit procedure (choice of the fit
range, shape of the threshold and background function).


The distribution of $M_{\mathrm{min}}$ for the $3\pi\pi^0\nu_\tau$ decay mode is shown
in  Fig. 4 together with the results of the fit with the same function.

The value of the $\tau$ mass estimator for this decay mode is
$p_1 = 1777.22 \pm 0.56$ MeV which is consistent within errors with
the result from the $\tau \rightarrow 3\pi^\pm \nu$ decay mode. 
As the statistical error for the $3\pi^\pm\pi^0\nu$ mode is significantly
larger than for the $3\pi^\pm \nu$ one, we will concentrate on the former 
decay mode only.

\section{Systematic uncertainties}
The following contributions to the overall systematic uncertainty
were considered:
%
%

\begin{figure}[htb]
\includegraphics[width=0.54\textwidth]{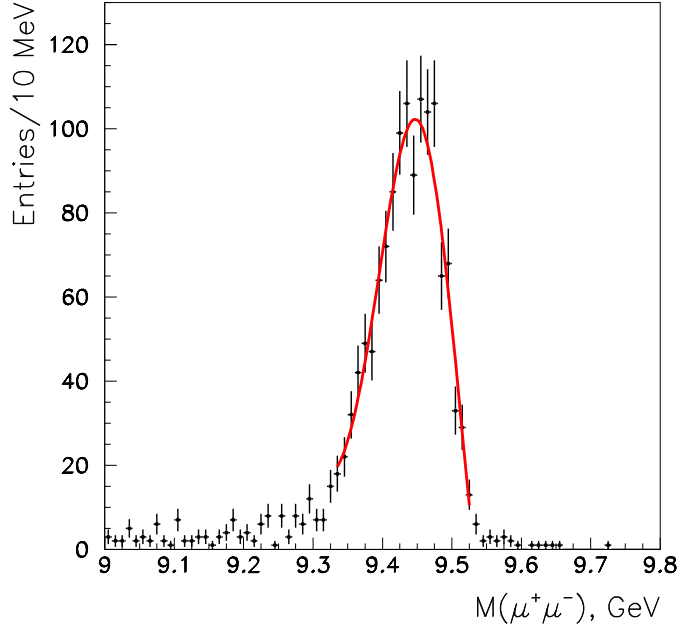}
\caption{The invariant mass distribution 
for $\Upsilon(1S)\rightarrow \mu^+\mu^-$} 
\label{fg05}
\end{figure}

\begin{itemize}

\item Calibration of the tracking system. We used muons
from the decay of $\Upsilon(1S)$, which are decay products of $\Upsilon(2S,3S)$
to $\Upsilon(1S)\pi^+\pi^-$. The peak position of $\Upsilon(1S)$ is
shifted from the nominal PDG value by $-4.5\pm2.3$ MeV.

The invariant mass distribution of $\mu^+\mu^-$ for the $\Upsilon(1S)$
peak is shown in Fig. 5. As the $\Upsilon(1S)$ is produced almost at rest
we use the relative mass shift of the visible peak position
from the PDG value as an estimate of the accuracy of the tracking
calibration, i.e. $\Delta M_\Upsilon / M_\Upsilon = \Delta P_\mu /P_\mu$.
The next assumption is that the relative shift of the momentum of 
the three pion system is equal to the relative shift of the single track
momentum, i.e. $\Delta P_{3\pi} /P_{3\pi} = \Delta P_\mu /P_\mu$.
This allows us to propagate the shift of $P_{3\pi}$ to a shift
of $M_{\mathrm{min}}$. 
In order to estimate the shift in the threshold position we finally take
$P_{3\pi} \approx P_\tau$ and $M_{\mathrm{min}} \approx M_\tau$,
because this is the region that is sensitive to the $\tau$ mass.
According to the above arguments, the relative shift of the $\Upsilon(1S)$
mass, which is equal to 4.5/9460.3 gives a systematic error from tracking
of 0.39 MeV.  

\item Choice of the fit range and the shape of the threshold function
for the $\tau$ mass estimation. In addition to the function 
$\mbox{arctan}((x-p_1)/p_2)$,
where $p_1$ and $p_2$ stand for the mass position and resolution respectively, 
we also tried the parametrizations $(x-p_1)/\sqrt{p_2+(x-p_1)^2}$ 
and $1/(1+exp((x-p_1)/p_2))$. 
%
To esimate the value of the correction to our estimator of the $\tau$ mass
and its uncertainty,
we used a Monte Carlo $\tau^+\tau^-$
sample with a statistics of approximately one half compared to the data
sample and an input $\tau$ mass of 1777.0 MeV.
We performed fits to the data and MC samples using three different
fit functions mentioned above in five ranges of $M_{\mathrm{min}}$.


From the fit to the MC distribution by Eq. (4) we get a 
difference between the visible threshold position
and the true input $\tau$ mass of $\delta p_1 = 0.70$ MeV.
The variation of the fit ranges and threshold function gives a
variation of the $\delta p_1$ within a $\pm$ 0.40 MeV range.

We correct the estimator value $p_1$ obtained from the fit by equation (4) 
to the data by
0.70 MeV and take the value of 0.40 MeV as an estimate of the
systematic uncertainty.


\item Uncertainty in the beam energy. 

For the estimation of this uncertainty we used the internal Belle
analyses of the full reconstructed $B$ decays for the 
energy calibration. In these analyses the reconstructed
$B$ meson energies were compared with the beam energies supplied by KEKB. 
The conclusion
from these analyses is that the beam energy is known with accuracy better than  
1.5 MeV.
This
uncertainty can be translated to an uncertainty in the $\tau$ mass from the  
following formula  
\begin{equation}
 \sigma(M_{\mathrm{min}})=\frac{E_X-P_X}{M_{\mathrm{min}}}\sigma(E_{\mathrm{beam}}).
\end{equation}

Near the threshold in the $M_{\mathrm{min}}$ distribution we can set 
$E_X \approx E_{\mathrm{beam}}$, $M_X \approx M_\tau$ and 
$M_{\mathrm{min}} \approx M_\tau$ which gives $\sigma(M) \approx 0.17 
\sigma(E_{\mathrm{beam}})$.
For $\sigma(E_{\mathrm{beam}})=1.5$ MeV we find $\sigma(M_\tau) \approx 0.26$ MeV.

To check the relation $\sigma(M) \approx 0.17 \sigma(E_{\mathrm{beam}})$,
we performed a simulation of $\tau$ decays with different $E_{\mathrm{beam}}$
values. We repeated the fit procedure for all MC samples assuming 
$E_{\mathrm{beam}} = M (\Upsilon (4S))/2$ and plotted the fit results versus the beam
energy used in the simulations in Fig. 6. 
The result of the straight line fit to the plot in Fig. 6 gives the value
of the slope of $P_2=0.1753\pm0.0002$, which is consistent with the analytical
calculation. 

\begin{figure}[htb]
\includegraphics[width=0.54\textwidth]{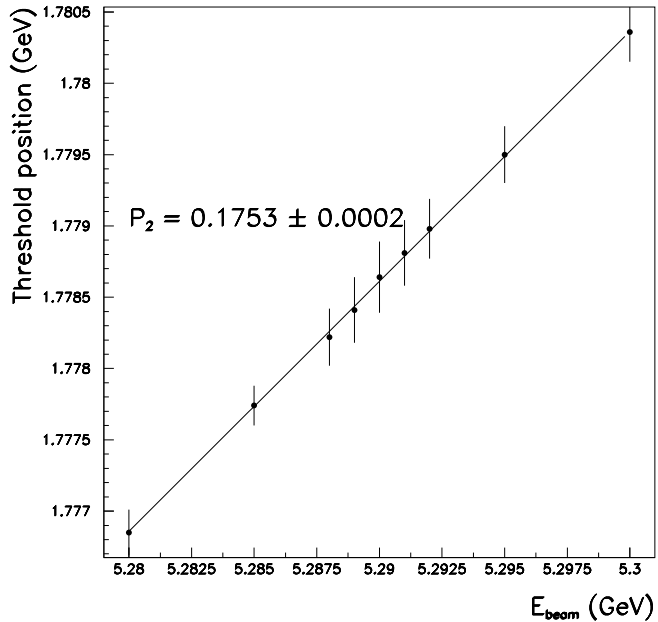}
\caption{The dependence of the fit value of the threshold position
on the beam energy used for the event simulation.
The line is a result of a straight line fit to this dependence.}
\label{fg06}
\end{figure}
\begin{figure}[htb]
\includegraphics[width=0.54\textwidth]{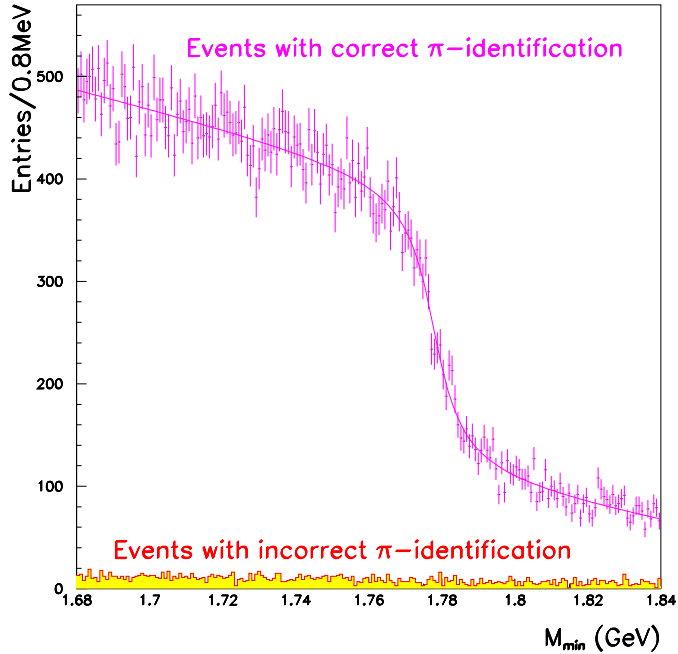}
\caption{The MC distribution of the pseudomass $M_{\mathrm{min}}$ for
the decays $\tau \rightarrow 3\pi^\pm \nu$ with correctly identified pions.
The filled histogram shows the contribution from incorrectly identified
particles.}  
\label{fg07}
\end{figure}

\item Systematic uncertainties coming from misidentified $\tau$
decays are negligible, since their $M_{\mathrm{min}}$ distributions show no
significant structure in the region of the $\tau$ mass. In Fig. 7
the MC $M_{\mathrm{min}}$ distributions are shown for correctly identified pions
and incorrectly identified particles. These distributions are obtained
from the same MC sample.
The background from non-$\tau^+\tau^-$ events can also be neglected.

\end{itemize}

Adding all these uncertainties in quadrature results in a total systematic
error of 0.62 MeV.
The final result is  
$M_\tau = 1776.71 \pm 0.25\mbox{(stat)} \pm 0.62\mbox{(syst)} \mbox{MeV}.$

\section{CPT test}
 The pseudomass method allows a separate measurement of the
masses of the positively and negatively charged $\tau$ leptons.
  
 A mass difference between positive and negative $\tau$ leptons would 
result in a difference in the energy between the $\tau$'s produced in the 
$e^+e^-$ collision.
This in principle makes the assumption $E_\tau = E_{\mathrm{beam}}$ invalid.
The distributions of the $M_{\mathrm{min}}$ for positive and negative $\tau$'s
decaying into $3\pi\nu$ are shown in Fig. 8 together with the results
of the fit. 

\begin{figure}[htb]
\includegraphics[width=0.54\textwidth]{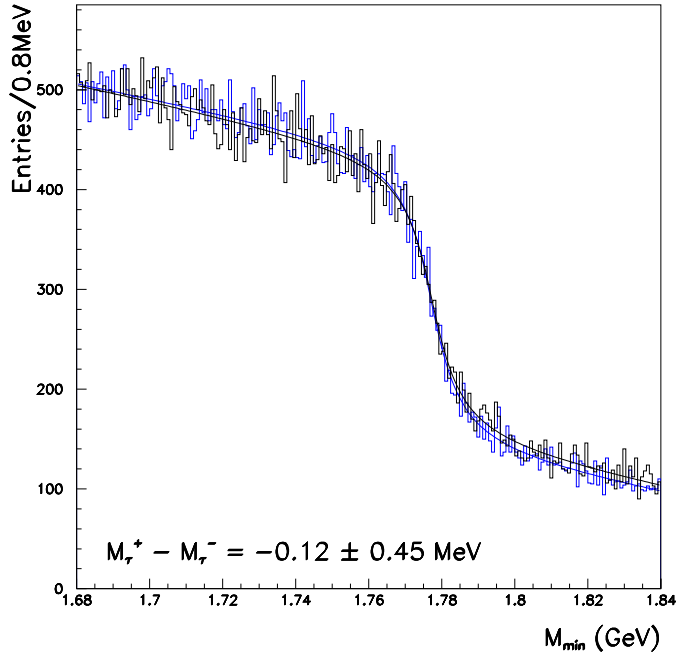}
\caption{The distributions of the pseudomass $M_{\mathrm{min}}$ for
the decays $\tau \rightarrow 3\pi^\pm \nu$ separately for
positive and negative $\tau$ decays.
In blue the distribution for positive, in black for negative $\tau$ decays is
shown.
The solid lines are the results of the fit with function (4). }
\label{fg08}
\end{figure}

Good agreement between the distributions for $\tau^+$
and $\tau^-$ is seen. The mass difference obtained from the independent fits
to these distributions is $M_{\tau^+}-M_{\tau^-} = -0.12\pm 0.45$ MeV.

Most sources of systematic errors affect the result for positive and 
negative $\tau$ leptons in the same way, so that their contributions
to the mass difference cancel. However, particles and antiparticles
interact differently with the detector material. 

To estimate a systematic shift
in the measurement of particle and antiparticle momenta we compared the
peak positions of $D^0 \rightarrow K^-\pi^+$ and 
$\bar{D^0} \rightarrow K^+\pi^-$, 
$\Lambda_c \rightarrow pK^-\pi^+$ and 
$\bar{\Lambda_c} \rightarrow \bar{p}K^+\pi^-$,    
$D^+ \rightarrow \phi(1020)\pi^+$ and $D^- \rightarrow \phi(1020)\pi^-$,
$D_S \rightarrow \phi(1020)\pi^+$ and $\bar{D_S} \rightarrow \phi(1020)\pi^-$.


The average relative mass shift from the
decay modes listed above is about $0.8 \times 10^{-4}$, which gives a systematic 
uncertainty in the mass difference between $\tau^+$ and $\tau^-$ of 0.15 MeV.

Adding the statistical and systematic errors in quadrature we obtain
$M_{\tau^+}-M_{\tau^-} = -0.12\pm 0.47$ MeV.

This result can be expressed as an upper limit on the relative mass 
difference~\cite{FC}

\begin{equation}
|(M_{\tau^+}-M_{\tau^-})|/M_{\mathrm{average}} < 5.0 \times 10^{-4} ~~ 
\mbox{at} ~90\%  ~\mbox{CL}. 
\end{equation}
 
 Without assuming CPT invariance it is no longer obvious that
the charges and masses of positive and negative $\tau$ decay products
should be the same. 
Good agreement of the $M_{\mathrm{min}}$ distributions  
for positive and negative $\tau \rightarrow 3\pi\nu$ decays shows that 
at the present
level of experimental accuracy CPT invariance is respected.

\section{Results}
 We have measured the mass of the $\tau$ lepton from the pseudomass
distributions of $\tau$ decays into three charged pions and neutrino.
The result is

\begin{equation}
M_\tau = 1776.71 \pm 0.25\mbox{(stat)} \pm 0.62\mbox{(syst)} ~\mbox{MeV}.
\end{equation}

 We obtained an independent measurement of the positive and 
negative $\tau$ mass. The measured values are consistent and an upper limit on
the relative mass difference is
$|(M_{\tau^+}-M_{\tau^-})|/M_{\mathrm{average}}$ is 
$5.0 \times 10^{-4} ~~ \mbox{at} ~90\%$ CL. 

\section{Acknowledgement}
We thank the KEKB group for the excellent operation of the
accelerator, the KEK cryogenics group for the efficient
operation of the solenoid, and the KEK computer group and
the National Institute of Informatics for valuable computing
and Super-SINET network support. We acknowledge support from
the Ministry of Education, Culture, Sports, Science, and
Technology of Japan and the Japan Society for the Promotion
of Science; the Australian Research Council and the
Australian Department of Education, Science and Training;
the National Science Foundation of China under contract
No.~10175071; the Department of Science and Technology of
India; the BK21 program of the Ministry of Education of
Korea and the CHEP SRC program of the Korea Science and
Engineering Foundation; the Polish State Committee for
Scientific Research under contract No.~2P03B 01324; the
Ministry of Science and Technology of the Russian
Federation; the Ministry of Higher Education, 
Science and Technology of the Republic of Slovenia;  
the Swiss National Science Foundation; the National Science Council and
the Ministry of Education of Taiwan; and the U.S.\
Department of Energy.

%
%

\end{document}